\documentstyle[12pt,titlepage,fleqn]{article}

\newcommand{\be}{\begin{equation}}
\newcommand{\ee}{\end{equation}}
\newcommand{\ba}{\begin{eqnarray}}
\newcommand{\ea}{\end{eqnarray}}

\newcommand{\non}{\nonumber}
\newcommand{\oh}{\frac{1}{2}}
\newcommand{\IIA}{type IIA superstring}
\newcommand{\IIB}{type IIB superstring}
\newcommand{\II}{type II superstring}
\newcommand{\het}{heterotic string}
\newcommand{\DD}{{\cal D}_L}
\newcommand{\DDA}{{\cal D}_{L1}}
\newcommand{\DDB}{{\cal D}_{L2}}

\begin{document}

\begin{titlepage}
\vspace*{0.1truecm}
\hfill{ hep-th/9604110}

\begin{center}

{\large\bf T-duality and Gauge Symmetry in Supermembrane Theory }

\vspace{1cm}

{\large Fermin ALDABE\footnote{E-mail: faldabe@phys.ualberta.ca}}

\vspace{1cm}

{\large\em Theoretical Physics Institute,
University of Alberta\\
Edmonton, Alberta, Canada, T6G 2J1}

\vspace{.5in}

\today \\

\vspace{1cm}
{\bf ABSTRACT}\\
\begin{quotation}
\noindent
T-duality has been shown to arise from a gauge symmetry in some
string theories.
However, in the case of \IIA\ compactified on a circle,
it is not possible to show that this is the case.
This situation is rather uncomfortable since string string duality
suggests that all
T-dualities should arise from a gauge symmetry.
Here we show that the T-duality of \IIA\ compactified on
a circle arises from the reparametrization of the supermembrane.
Then we show how this reparametrization can be understood in terms
of a gauge symmetry of the supermembrane, thus allowing to connect
T-duality in \II\ to a gauge symmetry of the supermembrane.

\end{quotation}
\end{center}

\end{titlepage}

\section{Introduction}

Not long ago, it was argued that some of the T-dualities 
in string theory can be 
interpreted as gauge symmetries \cite{DS,HP}.  This is the case of the T-duality
found in the bosonic and heterotic strings compactified on a circle.  On the
other hand, it does not seem possible to interpret T-duality as a gauge symmetry
in all cases.  For example, \IIA\ compactified on a circle of radius $R$ has 
as T-dual the 
\IIB\ compactified on a circle of radius $2/R$ \cite{DS}.  However,
there is no  enhanced gauge symmetry available at the self
dual radius which is needed to interpret the T-duality as a gauge symmetry.
This obstacle should be overcommed if we are going to take string string duality
and string unification seriously.  It would be strange to find
that T-duality in heterotic string theory which can be interpreted as a gauge
symmetry suddenly looses this property after a string-string duality 
transformation.
It is not natural to have  T-duality in heterotic 
string arise from
a gauge symmetry and not see the same thing happening in the \II; specially
when we have evidence that both theories are 
phases of the same M-theory.

There are three reasons which
tempt us to ask if it is possible that the T-dualities 
of \II\ compactified on a circle follow from a 
 gauge symmetry of the membrane action.
First, the interaction of both \II \ theories compactified on a circle
 are equivalent \cite{DS}. Second,
string theory can be obtained from membrane theory after dimensionally reducing
both the world volume and spacetime
\cite{D}. Third,
membrane theory is conjectured to be non 
renormalizable because 
it contains both perturbative and nonperturbative effects of
string theory \cite{T}\footnote{This was also argued in \cite{EW} where it was
shown that \IIA\ in the strong coupling limit yields 
11-dimensional supergravity.}; thus, all string interactions are included in the
membrane action.
The natural setting to answer this question was provided in \cite{Sc} where
it was shown, examining the non perturbative spectrum,
that \IIB\ and \IIA, both compactified on a circle
 should be identified 
with 11 dimensional supergravity compactified on a torus. 
Since 11 dimensional supergravity is believed to be 
the low energy limit of membrane theory
\cite{TBS,yo1} we should consider the membrane compactified on a torus to 
show that the T-duality of \II\ compactified on a circle arises from a
reparametrization invariance of the membrane which can
be shown to follow from a gauge symmetry present in the supermembrane action.  
This result provides more
evidence on the conjecture that the different string theories are
different phases of the supermembrane.  It also provides a further step in
the unification of T-dualities which is essential in the understanding of
string-string duality. 

The strategy used in \cite{DS} to show that T-duality in the \het\
is a gauge symmetry consists in showing that 
at the enhanced symmetry point 
the action of T-duality on the vertex operator of the
gauge field can be interpreted as a rotation of the SU(2)
gauge group about the ``x-axis" of the SU(2), and is therefore a
gauge symmetry.  Despite the demonstration taking place at the self dual radius,
the duality transformation is a gauge transformation everywhere in the moduli 
space.  In the \II\ there are no points in the moduli space at which
the theory acquires additional gauge symmetry, and therefore, it is not
possible to show that T-duality is a gauge symmetry of this theory.

We shall show that T-duality in \II\ is a reparametrization 
of the supermembrane.    We will then consider a particular point in the moduli 
space of the membrane where the U(1) symmetry is enhanced to an SU(2) symmetry.
This reparametrization of the membrane at the enhanced symmetry point, 
which is interpreted as a T-duality transformation in string theory,
may then be interpreted as a spacetime gauge symmetry: it can be treated as an
SU(2) rotation by $\pi$ about the ``x-axis".
It would be interesting to see how our work is related to that of 
\cite{O} which also deals with the interpretation of T-dualities in M-theory.

The action for the membrane \cite{TBS} is given by
\be
S=\int d^3\zeta (\oh \sqrt{-g}g^{ij}E^a_iE^b_j\eta_{ab}-\frac{1}{6}
\epsilon^{ijk}E^A_iE^B_jE^C_kA_{CBA}-\oh\sqrt{-g}).\label{s}
\ee
Here, $E_i^A$ is a supervielbein and $A_{CBA}$ is a super three-form.
The world volume  has coordinates $\zeta$ and metric $g_{ij}$.
Among the symmetries of this action  there is 
local reparametrization symmetry.   The presence of a fermionic $\kappa$
symmetry and supersymmetry will allow us to concentrate on the  bosonic sector.
The bosonic sector of action (\ref{s}) is given by
\be
S=\int d^3\zeta (\oh \sqrt{-g}g^{ij}\partial_ix^m\partial_jx^nG_{mn}
-\frac{1}{6} \epsilon^{ijk}\partial_ix^m\partial_jx^n\partial_kx^pA_{pnm}
-\oh\sqrt{-g}).\label{s1}
\ee
The $x^m,\ m=1,...,11$ are coordinates on an 11 dimensional 
 manifold with metric $G_{mn}$.
As suggested in \cite{Sc}, the supermembrane compactified on a torus should
be equivalent to \II\ string compactified on a circle which suggest that 
two of the coordinates should be coordinates on a torus
and thus have periodic boundary conditions
\be
x^I=x^I+2\pi R^I, \ \ I=10,11
\ee
The moduli space of the torus is characterized by complex and Kahler structure
deformations.  It is then possible to pick a point in the moduli space of the
torus such that
\be
R^{11}R^{10}=2.\label{a4}
\ee

In order to obtain a string theory from action (\ref{s}) we must dimensionally
reduce the action in the same way as was done in \cite{D}.  First we
split the world volume  coordinates
\ba
\zeta^i&=&z^i\ \ i=1,2\non\\
\zeta^3&=&\rho.
\ea
We then may use the reparametrization invariance to set
\be
\rho=x^{10}\label{x10}
\ee
or to set
\be
\rho=x^{11}\label{x11}
\ee

Standard techniques in dimensional reduction \cite{Sc0} suggest we write the
world volume metric  in the form
\be
g^{ij}=\phi^{-2/3}\pmatrix{
\hat{g}^{ij}+\phi V_iV_j & \phi^2 V_i \cr
\phi^2 V_j & \phi^2 \cr
}.
\ee
The fields $\phi$ and $V$  are non dynamical.
Making use of the equations of motion and the choice
(\ref{x10}) we obtain a string action of the form
\be
S=\int d^2z  \sqrt{-\hat{g}}\hat{g}^{ij}
\partial_i\hat{x}^m\partial_j\hat{x}^n\hat{G}_{mn}+\frac{1}{2}
\epsilon^{ij}\partial_i\hat{x}^m\partial_j\hat{x}^n\hat{A}_{nm},\;\;\;
 i,j=1,2. \label{s2}
\ee
The $\hat{x}^m,\ \ m=1,...,10$ are now coordinates on an 10 dimensional 
manifold with metric $\hat{G}_{mn}$. 
The action (\ref{s2}) is the Green-Schwarz
action \cite{GS} for the \II\ propagating on a circle of radius 
$R^{10}$.  If we use the reparametrization invariance (\ref{x11}) we obtain
the same action but the string now propagates on a circle of radius
$2/R^{10}$.  Thus, the T-duality found in \II\ on a circle is just a certain
reparametrization of the supermembrane.
This result was obtained at a particular point on the moduli space of the
membrane theory.  Similar arguments to those used in \cite{DS} allow us
to extend our result to all the points in the moduli space.  Thus,
this reparametrization of the supermembrane compactified on
a torus is responsible for the T-duality found in the \II\ compactified on a
circle.  

The fact that a reparametrization of the membrane is responsible for T-duality
in the type II string  should not come as a surprise.  We know of other
T-dualities in nine dimensions where the heterotic string is related to the
type I string \cite{HW}.  It turns out that this T-duality can be interpreted
as a reparametrization of the open membrane which have current algebras
defined at its boundaries \cite{yo1}.  As the open membrane which has the
topology of the cylinder reduces its length, it becomes a heterotic string.
As the open membrane reduces its radius, it becomes an open string, and
the current algebra generates the Chan Paton factors required to define
the type IA string.

We have succeeded in relating T-duality in \II\  to a certain reparametrization 
invariance of the supermembrane.  Now, we must relate this certain 
reparametrization invariance to a  gauge symmetry.  To do this we first
note that for a compactification of the supermembrane on a manifold
which has $n$ 2-cycles, the theory will have $n$ gauge fields \cite{yo}.  
To see this
it is sufficient to consider the term 
\be
-\int d^3\zeta
\frac{1}{6} \epsilon^{ijk}\partial_ix^m\partial_jx^n\partial_kx^pA_{pnm}
\label{tres}
\ee
in action (\ref{s1}).  As it is done in string theory \cite{RW}, we 
may express 
\be
A_{11\ 10 m}=A_{11\ 10}^I A^I_m\label{2c}
\ee
where $I$ labels a particular 2-cocycle (for the torus $I$=1), and
$m$ labels a spacetime coordinate 
while $10$ and $11$ label compact coordinates.  In order
argue for the existence of nonperturbative states which will enhance this
U(1) symmetry to an SU(2) symmetry it is best to digress first to
a K3 compactification of the supermembrane \cite{yo}.  

As it was explained in 
\cite{T2,HT}, in order to have D=7 membrane/string duality, it is necessary
to consider the nonperturbative states (solitonic membranes) which can
wrap about the 2-cycles of K3.  As a 2-cycle of K3 becomes massless it will
enhance the U(1) symmetry associated to that 2-cycle to an SU(2) symmetry.
The U(1) gauge fields in the supermembrane are constructed from the 3-form
tensor $A_{abm}=A_{ab}^I A^I_m$ where $I$ denotes a particular K3 two-cocycle
and $a,b$ label compact coordinates while $m$ denotes  spacetime coordinates.
The states which enhance the U(1) symmetries are given by the nonperturbative
states whose masses are proportional to the area of the 2-cycles.

Membrane/string duality requires this enhancement of symmetry
to hold for all of the 22 cycles which vanish at
particular points in the moduli space.  In particular, this means that
it must also hold for any of 
the 22 cycles of $T^4/Z_2$, one of  which 
is the  2-cycle of a $T^2$ which is left invariant under the action of the
orbifold.  Thus, if the two-cocycle $A_{10\ 11}$ in (\ref{2c}) which is
dual to the 2-cycle of $T^2$
vanishes, 
two charged states
with respect to the U(1) field will become massless and enhance the symmetry
to an SU(2) gauge group.  The symmetry enhancement is independent of how 
the size of the 2-cycle vanishes.  In particular, we can consider the
case in which the parallelogram describing the two-cycle reduces to a line.
This is achieved by reducing either $x^{10}$ or $x^{11}$ to zero size.
Thus, as $x^{10}$ or $x^{11}$ vanishes to zero size, the gauge symmetry is
enhanced to SU(2).

Now consider the case in which we set 
\be
\rho=x^{11}\label{o1}
\ee
Then (\ref{tres}) reduces to
\be
-\int d^3\zeta
\frac{1}{6} \epsilon^{ij}\partial_ix^m\partial_jx^{10}A_{1011m}
\label{tres1}
\ee
where $i,j=1,2$.  
We now use the reparametrization invariance to set
\be
\rho=x^{10}\label{o2}
\ee
Then (\ref{tres}) reduces to
\be
-\int d^3\zeta
\frac{1}{6} \epsilon^{ij}\partial_ix^m\partial_jx^{11}(-A_{1011m})
\label{tres2}
\ee
Thus, the reparametrization that takes (\ref{o1}) to  (\ref{o2}) has
the effect of interchanging $x^{11}$ with $x^{10}$ and therefore of taking
\be
A_m^I\to -A^I_m\label{rr}
\ee
which is also what it is expected from a T-duality transformation in string
theory.  In performing this reparametrization when either $x^{10}$ or
$x^{11}$ is very small, we are  actually enforcing (\ref{rr}) at an 
enhanced symmetry point.  Therefore, the reparametrization invariance
which takes (\ref{o1}) to (\ref{o2}) is a gauge symmetry because it can
be understood as a rotation about the SU(2) 
``x-axis" by $\pi$, just like T-duality
is understood as a gauge symmetry in the bosonic and heterotic strings 
\cite{DS}.

This result can be extended
to \II\ theories compactified in more dimensions,  for example, when 
compactifying \II\ theories on a torus.  
In this case we may classify theories according to the radii of compactification
and to whether they are \IIA\ or \IIB.  T-duality in this case can act
on either of the radii of the torus.  Because T-duality changes the sign
of the fermionic partners of the compactified dimensions, an odd number of
T-duality transformations will change the sing of $\Gamma^{11}$ and thus
alter the GSO projection \cite{DS}.  Thus, under an odd number of duality
transformations the \IIA\ is mapped to the \IIB\ and vice versa.  
On the other hand, an even number
of T-duality transformations maps the \IIA\ (\IIB\ ) to itself.  In all cases,
the radius on which T-duality acts is mapped to its dual.  Denoting $\DD$ 
the T-duality transformation we have the following maps
\ba
\DDA|IIA, R_1,R_2>&=& |IIB,2/ R_1,R_2>\\
\DDB|IIA, R_1,R_2>&=& |IIB, R_1,2/R_2>\\
\DDA\DDB|IIA, R_1,R_2>&=& |IIA,2/ R_1,2/R_2>\\
\DDA\DDB|IIB, R_1,R_2>&=& |IIB,2/ R_1,2/R_2>
\ea
The first two maps
relate three different theories which
 should be identified because they
have similar spectra and 
interactions \cite{DS}.  On the other
hand, the last two maps must be used to  identify same
theories.
These maps are not gauge symmetries of the \II\ theories. 
However, 
the T-duality maps of the \II\ are a consequence of reparametrizations of the
supermembrane.  These
reparametrizations of the supermembrane can take place at
the points in the moduli space where one of the three 2-cycles
of $T^3$ vanishes to enhance one of the three U(1) gauge symmetries to an SU(2)
gauge symmetry.   Thus, these reparametrizations can also be identified with
gauge transformations.
Let us denote the radii of the three-torus $R^{11},\ R^{10},
\ R^{9}$.  We shall choose the radii to satisfy 
\be
R^{11}=\frac{2}{R^{10}}=\frac{2}{R^{9}}.
\ee
Using  the reparametrization invariance of the supermembrane to set
\be
\rho=x^{9}\label{y9}
\ee
or 
\be
\rho=x^{10}\label{y10}
\ee
or 
\be
\rho=x^{11}\label{y11}
\ee
we obtain the \II\ theories
\ba
|IIA,R^{10},R^{11}>,\\
|IIB,2/R^{10},R^{11}>,\\
|IIB,R^{10},2/R^{11}>,
\ea
respectively.  Here we have adopted the convention that the gauge
choice (\ref{y9}) yields the $|IIA,R^{10},R^{11}>$.
Thus, once more, the T-dualities which do not seem to be a symmetry of the
\II\ happen to follow from the reparametrization of the supermembrane.
These reparametrizations can be understood as gauge symmetries, 
just like in the case of the $T^2$
compactification of the supermembrane.

One of the problems we have encountered in the past is the existence of 
certain T-dualities which arise  from gauge symmetries and other T-dualities
which do not. This is an obstacle to our understanding of string unification.
However, it is believed that string theory is a phase
of supermembrane theory and 
the proof that the T-dualities which do not
seem to be a consequence of gauge symmetry in string theory,
arise in fact from the 
gauge symmetry of the membrane add more evidence to the conjecture
that membrane theory will unify all string theories.

\pagebreak

\end{document}